\documentclass[aip,jmp,amsmath,amssymb,reprint,floatfix]{revtex4-1}
\usepackage{braket}
\usepackage{graphicx}
\usepackage{xcolor}
\usepackage{hyperref}

\begin{document}
\title{Spin in the Extended Electron Model}
\author{Thomas Pope}
\email{Thomas.Pope2@newcastle.ac.uk}
\author{Werner Hofer}
\affiliation{School of Chemistry, Newcastle University, Newcastle NE1 7RU, 
United Kingdom}
\date{\today}
\begin{abstract}
 It has been found that a model of extended electrons is more suited to 
 describe theoretical simulations and experimental results obtained via 
 scanning tunnelling microscopes, but while the dynamic properties are easily 
 incorporated, magnetic properties, and in particular electron spin properties 
 pose a problem due to their conceived isotropy in the absence of measurement. 
 The spin of an electron reacts with a magnetic field and thus has the 
 properties of a vector. However, electron spin is also isotropic, suggesting 
 that it does not have the properties of a vector. This central conflict in 
 the description of an electron's spin, we believe, is the root of many of the 
 paradoxical properties measured and postulated for quantum spin particles. 
 Exploiting a model in which the electron spin is described consistently in 
 real three-dimensional space - an extended electron model - we demonstrate 
 that spin may be described by a vector and still maintain its isotropy. In 
 this framework, we re-evaluate the Stern-Gerlach experiments, the Einstein-
 Podolsky-Rosen experiments, and the effect of consecutive measurements and 
 find in all cases a fairly intuitive explanation. 
\end{abstract}
\maketitle

\section{Introduction}
 Magnetic fields are the manifestation of charge in rotation around a 
 centre~\cite{jackson1999classical}. In single atoms, the orbit of electrons 
 around a nucleus is accounted for by the so-called orbital magnetic dipole 
 moment~\cite{eisberg1986quantum,mahajan1988electricity}. This describes a 
 magnetic field with a well know magnitude and orientation. However, it has 
 been observed that atoms with no orbital magnetic moments, like silver, 
 experience a force upon application of an external magnetic 
 field~\cite{gerlach1922experimentelle}, which has been attributed to the spin 
 of the atom's outer electron(s). Here, the classical picture of magnetic 
 moments breaks down and, in the standard model, we conclude that electron 
 spin is not an object in real space because it is isotropic and therefore 
 does not have the properties of a vector. 
 
 In the standard model, electrons are modelled as point 
 particles~\cite{burkhardt2008foundations}, which is, we believe, the 
 fundamental problem with conventional interpretations. With this restriction, 
 the only way to reconcile that electron spin is isotropic in one case and 
 vector-like in another is to rely on abstract mathematics. In addition, 
 recent experimental evidence cast further doubt on this 
 assumption~\cite{rieder2004scanning}, since current STM measurements appear 
 able to resolve a density distribution on noble metal surfaces that cannot be 
 explained as a consequence of a probability distribution of detection events 
 without violating Heisenberg's uncertainty 
 relations~\cite{hofer2012heisenberg}. If, instead, we relax this condition 
 and employ an extended electron 
 model~\cite{hofer2011unconventional,hofer2014elements}, it is possible to 
 render these two properties of electron spin in real space. This extended 
 electron model is based on four postulates. Firstly, the wave properties of 
 electron are a \emph{real} property of electrons in motion. This accounts for 
 the high resolution in STM experiments. Secondly, electrons in motion possess 
 intrinsic electromagnetic potentials and, thirdly, these give rise to the 
 intrinsic magnetic moment, or spin, of electrons. Finally, in equilibrium, 
 the energy density is constant throughout the space occupied by an electron. 
 Within this framework, formulated using geometric 
 algebra~\cite{hestenes2012clifford,gull1993imaginary}, it is possible to 
 characterise an electron spin vector in real three-dimensional space while 
 reproducing the results of experiment and maintaining isotropy. 
 
 In the following, we present the standard approach along with the extended 
 electron approach and discuss their implications with regard to experimental 
 results. 
 
\section{Pauli Algebra}
 Within the standard model, spin is accounted for by the Pauli matrices, which, 
 along with the identity matrix, form a complete basis for all $2\times2$ 
 Hermitian matrices and, as observables correspond to Hermitian operators, 
 they span the space of observables of the 2-dimensional Hilbert 
 space~\cite{benenti2004principles},
\begin{align}
 \sigma_{x}&=\ket{0}\!\!\bra{1}+\ket{1}\!\!\bra{0},\nonumber\\
 i\sigma_{y}&=\ket{0}\!\!\bra{1}-\ket{1}\!\!\bra{0},\nonumber\\
 \sigma_{z}&=\ket{0}\!\!\bra{0}-\ket{1}\!\!\bra{1}.
\end{align}

 Each matrix has eigenvalues of $\pm1$ representing spin-up and spin-down. In 
 the case of spin-1/2 particles, we define spin operators, 
 $\text{\bf{S}}_{a}=\hbar\sigma_{a}/2$, where $a$ is axis ($x$, $y$, or $z$) 
 and the corresponding eigenvectors in Hilbert space are given by,
\begin{align}
 \ket{x^{\pm}}&=\frac{1}{\sqrt{2}}\left(\ket{0}\pm\ket{1}\right),
 \text{\hspace{0.02\textwidth}}
 &\ket{z^{+}}&=\ket{0},\nonumber\\
 \ket{y^{\pm}}&=\frac{1}{\sqrt{2}}\left(\ket{0}\pm i\ket{1}\right),
 \text{\hspace{0.02\textwidth}}
 &\ket{z^{-}}&=\ket{1}.
\end{align}

 Generally, the quantum state of a particle, with respect to spin, is 
 represented by a two component spinor, 
 $\Psi=\varPsi_{0}\ket{0}+\varPsi_{1}\ket{1}$, which contains a superposition 
 of both states. When the spin of this particle is measured with respect to a 
 given axis, $a=x,y,z$, the probability that a spin of $\pm\hbar/2$ will be 
 measured is $\left|\braket{a^{\pm}|\Psi}\right|^{2}$. Following the 
 measurement, the spin state of the particle is said to collapse into the 
 corresponding eigenstate and all equivalent measurements will yield the same 
 eigenvalue, but when a measurement is performed on another axis, $b\neq a$, 
 the probability of finding a spin of $\pm\hbar/2$ is then $\left|
 \braket{b^{\pm}|a^{\pm}}\right|^{2}=1/2$. Going on to remeasure along the 
 original axis, we find we are equally likely measure either spin-up or spin-
 down, so there is no memory of the original measurement. Mathematically, this 
 is due to the non-commutativity of the Pauli matrices, $[\sigma_{b},
 \sigma_{a}]\neq0$. Physically, the explanation is not clear, which is a 
 consequence of our failure to define the physical process responsible for the 
 wavefunction collapse. 

 Describing this process has been problematic. Objective collapse theories 
 like the Ghirardi-Rimini-Weber theory~\cite{ghirardi1986unified} or the 
 Penrose interpretation~\cite{penrose1996gravity} adopt a more rigorous 
 version than the Copenhagen interpretation~\cite{heisenberg1958language}, but 
 these have been challenged experimentally~\cite{knee2016strict}.  Adopting de 
 Broglie's ontological approach~\cite{debroglie1925research}, as opposed to 
 Schr\"{o}dinger's more epistemological 
 approach~\cite{schrodinger1926undulatory}, one may argue that electron spin  
 can be described in real space. In the de Broglie-Bohm 
 model~\cite{bohm1952suggested,bohm1952suggestedII}, or in other hidden-
 variable approaches, we find non-local potentials due to Bell's 
 inequalities~\cite{bell1966problem}. Suggested solutions to this problem 
 include the many-worlds interpretation~\cite{everett1957relative}, 
 superdeterminism~\cite{hooft2007free,hooft2009entangled}, and 
 retrocausality~\cite{de1976time,dowe1997defense}, but these depend on a 
 somewhat profound metaphysical shift in our  description of the universe. We 
 note that many loopholes exist in the Bell's inequalities 
 experiments~\cite{santos2004failure}. Indeed, strictly speaking, no Bell 
 experiment can exclude all conceivable local hidden-variable 
 theories~\cite{brunner2014bell} and, as there is no physical reality ascribed 
 to the imaginary component of the phase of the two measured objects, the 
 description is limited from certain viewpoints~\cite{hofer2012solving}. This 
 will be explored in more detail in the following sections.

\section{Extended Electrons}
 In the extended electron model, we exploit the framework of geometric algebra 
 to parametrise electron spin in real space. Firstly, we define three 
 perpendicular unit vectors, ${\bf{e}}_{1}$, ${\bf{e}}_{2}$, and 
 ${\bf{e}}_{3}$. Correspondingly, we may define three perpendicular bivector 
 terms as the plane cast by each combination of two unit vectors, 
 ${\bf{e}}_{1}{\bf{e}}_{2}$, ${\bf{e}}_{2}{\bf{e}}_{3}$, and 
 ${\bf{e}}_{3}{\bf{e}}_{1}$. No plane is cast by two parallel vectors, so 
 ${\bf{e}}_{i}{\bf{e}}_{i}=1$. We then define the trivector, which corresponds 
 to the unit volume defined by the three unit vectors. This we call the 
 pseudoscalar, ${\bf{i}}={\bf{e}}_{1}{\bf{e}}_{2}{\bf{e}}_{3}$. Multiplying 
 the pseudoscalar with a vector gives  the bivector perpendicular to the 
 vector, ${\bf{i}}{\bf{e}}_{1}={\bf{e}}_{2}{\bf{e}}_{3}$. We note that the 
 behaviour of the Pauli matrices is implicitly reproduced by the elements of 
 geometric algebra~\cite{doran1993states}. Indeed, the Pauli matrices are a 
 matrix description of rotations in three dimensional space, described in 
 geometric algebra by the bivectors.

\begin{figure}[!ht]
 \centerline{\includegraphics[width=0.9\linewidth]{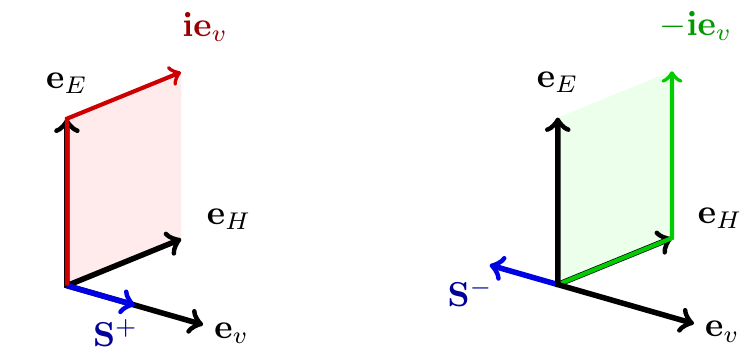}}
 \caption{Schematic of electron spin and field vectors (${\bf{S}}$, 
 ${\bf{e}}_{E}$, and ${\bf{e}}_{H}$,respectively) and the vector of motion, 
 ${\bf{e}}_{v}$, for an electron, with both the parallel (left) and 
 antiparallel (right) behaviour. The direction of the electron spin vector, 
 ${\bf{S}}^{\pm}$, is shown by the short arrows on the ${\bf{e}}_{v}$ axis in 
 both cases.}
 \label{field_vectors}
\end{figure}

 We now define a vector of motion, ${\bf{e}}_{v}$, and the bivector term 
 perpendicular to the vector of motion, ${\bf{i}}{\bf{e}}_{v}$. This bivector 
 term may be visualised as the product of two vectors, which are perpendicular 
 to one another and to the vector of motion, ${\bf{e}}_{E}$ and ${\bf{e}}_{H}$, 
 such that the bivector is given by 
 ${\bf{e}}_{E}{\bf{e}}_{H}={\bf{i}}{\bf{e}}_{v}$ (see 
 figure~\ref{field_vectors}). These vectors correspond respectively to the 
 direction of the electric, $\text{\bf{E}}$, and magnetic, $\text{\bf{H}}$, 
 field, which we introduce in accordance with the second postulate of the 
 extended electron model. An additional phase is also added to account for the 
 energy conservation of electrons at the local 
 level~\cite{hofer2011unconventional}, but for simplicity we set this term to 
 zero in our notation.
 
 Since the geometric product is anti-commutative, we may also define an 
 antiparallel bivector, ${\bf{e}}_{H}{\bf{e}}_{E}=-{\bf{i}}{\bf{e}}_{v}$. Thus, 
 this bivector term gives rise to a spin vector, with a direction 
 corresponding to the spin unit vector, ${\bf{e}}_{S}$, that can be either 
 parallel or antiparallel to the vector of motion. The electron spin is 
 defined by the helictiy and relative direction of the electromagnetic field 
 terms, which satisfies the third postulate of the extended electron model. 
 Indeed, we may define the a Poynting-like vector, which we call the spin 
 density, $S=\left|\text{\bf{E}}\right|\left|\text{\bf{H}}\right|$, that gives 
 the energy density of the field components of the electron. In this framework, 
 the wavefunction may be written in terms of the mass density, $\rho$, the 
 spin density, and the direction of the spin vector,
 \begin{equation}
  \Psi\left({\bf{r}}\right)=\rho^{1/2}\left({\bf{r}}\right)+{\bf{i}}
  {\bf{e}}_{S}\left({\bf{r}}\right)S^{1/2}\left({\bf{r}}\right).
 \end{equation}
 Here, all terms depend on position, ${\bf{r}}$, which we omit from further 
 equations, but note that rather than describing a probabilistic distribution 
 of point-like electron states, this wavefunction describes the physical 
 properties within the volume of an extended electron.
 
 The duality operation, $\Psi^{\dagger}$, is represented by a change in the 
 helicity of the bivector term and, hence, a change in sign of the spin vector. 
 The product of $\Psi$ and $\Psi^{\dagger}$ complies with the Born rule and 
 corresponds to the inertial number density of the electron,
 \begin{equation}
  \Psi\Psi^{\dagger}=\rho+S=\rho_{0},
 \end{equation}
 which corresponds to the requirement of energy conservation and the fourth 
 postulate of the extended electron model; that the energy density at every 
 point of the extended electron is a constant. Here, the wave properties are 
 related to oscillations in the mass density of the electron, which are 
 supplemented by equal and opposite oscillations in the spin density, 
 $\dot{S}=-\dot{\rho}$. So the first postulate of the extended electron model 
 is satisfied.

 Figure 1 shows a schematic representation of these bivector 
 notation, where the spin vector of an electron is given 
 by~\cite{hofer2011unconventional},
 \begin{equation}
  {\bf{S}}^{\pm}=\frac{\hbar}{2}\Psi{\bf{e}}_{S}^{\pm}
  \Psi^{\dagger}=\pm\frac{\hbar}{2}\rho_{0}{\bf{e}}_{v},
 \end{equation}

 Electron spin, defined in this way, is a constant vector associated with the 
 direction perpendicular to the plane of the electromagnetic field terms, 
 which are defined by the velocity of the electron. Electrons with vanishing 
 velocity, therefore, contain no field components to their energy density and 
 thus do not possess spin. The spin of an electron in motion is only isotropic 
 in relation to rotations in the bivector plane perpendicular to the vector of 
 motion, ${\bf{i}}{\bf{e}}_{v}$, but since this direction is due to the motion 
 of the electron, a statistical manifold of equal number spin-up, 
 ${\bf{S}}^{+}$, and spin-down, ${\bf{S}}^{-}$, electrons is fully isotropic.
  
 If the electrons are free, a magnetic field, ${\bf{B}}$, will alter their 
 trajectory according to the classical Lorentz forces, but if, on the other 
 hand, they are not free - instead moving along a constrained trajectory - 
 their spin will be affected. This effect is modelled by a modified Landau-
 Lifshitz equation~\cite{hofer2011unconventional}, 
\begin{equation}
 {\bf{\dot{e}}}_{S}=\text{const}\cdot{\bf{e}}_{S}\times
 \left({\bf{u}}\times{\bf{\dot{B}}}\right),
\end{equation}
 where ${\bf{u}}$ is the electron's velocity.  For a finite static field, the 
 induced spin vector, ${\bf{S}}'$, may be described by the first order term,
 \begin{equation}
  {\bf{S}}'=\text{const}\cdot{\bf{S}}\times\left({\bf{u}}\times{\bf{B}}\right).
 \end{equation}
 So, in response to an external magnetic field, the spin vector rotates in 
 either a parallel or antiparallel direction depending on electron spin, which 
 gives rise to two induced spin densities. The induced spin densities will 
 lead to a precession around the magnetic field in two directions, which will 
 give rise to induced magnetic moments parallel, or anti parallel to the field. 
 In an inhomogeneous field the force of deflection is then directed either 
 parallel or antiparallel to the field gradient, leading to the alternate 
 trajectories seen in Stern-Gerlach-type 
 experiments~\cite{gerlach1922experimentelle}. For example, in the case of the 
 hydrogen atom, the wavefunction is an exponentially decaying wavefunction, 
 similar to that in the standard picture~\cite{burkhardt2008foundations}, but 
 with the electron spin direction parallel to the radial vector and pointing 
 outward (spin-up) or inward (spin-down). These spin components are acted upon 
 by the magnetic field and split the trajectory of the atoms accordingly. Here, 
 there is no wavefunction collapse, we simply reveal the direction of the spin 
 vector with respect to the vector of motion. The conventional framework omits 
 the possibility that measurements directly affect the electron spin 
 properties of a system. We see in the extended electron model that the 
 measurement has an explicit effect. Thus, we can explain why measurements on 
 different axes are non-commutative: the measurement is felt by the electron 
 and the spin vector is realigned with each new measurement. 

\section{Spooky action at a distance}
 We now consider the famous Einstein-Podolsky-Rosen (EPR) thought 
 experiment~\cite{einstein1935can,einstein1936physics}, which concludes that 
 communication between two measurements seems to violate the principle of 
 causality. We imagine a source that emits an electron pair. The spin of the 
 two particles are measured separately, but due to their common source, the 
 measurements implicitly depend on one another. If the z-axis of the first 
 electron is measured to be spin-up, then it is known that the z-axis 
 measurement of the second electron will be spin-down. In the standard model, 
 this is because the initial measurement has collapsed the 
 wavefunction~\cite{thaller2005advanced}. On the other hand, if the 
 measurement on the second electron is performed along the y-axis, there is an 
 equal chance of measure spin-up or spin-down. The implication is that the 
 second electron somehow \emph{knows} on which axis the first measurement was 
 performed, a phenomenon that Einstein dubbed spooky action at a distance. 
 Experimental evidence has thus far shown a correlation between these two 
 measurements~\cite{wittmann2012loophole}, but that on its own is not enough 
 to prove a causal link. In the framework of extended electrons, we find that 
 this communication is an epiphenomenon of an underlying 
 correlation~\cite{hofer2012solving}, which is contained mathematically in 
 local variables.

 The spin vector can either be parallel or antiparallel to the vector of 
 motion so if measurements are taken parallel to this axis for both electrons, 
 the correlation between the two measurement is explained trivially. Indeed, 
 this argument can be extended to all measurement angles except perfectly 
 perpendicular, at which point the probability of measuring spin-up or spin-
 down are equal and correlations between the measurements are harder to 
 explain. Here, we assume the measurement contains rotations in the plane 
 perpendicular to the direction of motion and the spin vector, which in 
 geometric algebra, are described by the multiplication of two vectors. The 
 term itself is given the name rotor. We describe a rotation on the plane 
 ${\bf{e}}_{E}{\bf{e}}_{H}$ through an angle of $\varphi$, by,
 \begin{equation}
  R(\varphi)=e^{\left({\bf{e}}_{E}{\bf{e}}_{H}\right)
  {\bf{e}}_{S}\varphi}=\cos{\varphi}+{\bf{i}}\sin{\varphi}.
 \end{equation}
 Then the probability of detecting an angle of rotation, $\varphi$, is given 
 by the square of the scaler part of the rotor, 
 \begin{equation}
  p\left(\varphi\right)=\cos^{2}{\varphi}.
 \end{equation}
 This is true regardless of whether electron spin is parallel or antiparallel, 
 since that effect is only apparent in the pseudoscalar term, 
 ${\bf{i}}\sin{\varphi}$, which changes sign from positive to negative 
 respectively. It is here that the model diverges from Bell's original 
 derivation of his inequalities, in which he assumes the correlation 
 probability is the product of the two measurement probabilities. Instead, to 
 account for the two rotations, we take the product of the rotors for each 
 electron, assuming that the latter spin is antiparallel,
 \begin{equation}
  R(\varphi_{1})\cdot R(\varphi_{2})=
  e^{{\bf{i}}\left(\varphi_{1}-\varphi_{2}\right)}.
 \end{equation}
 The square of the scaler term then gives the correlation probability in a 
 form similar to that derived in the Clauser-Horne-Shimony-Holt 
 formalism~\cite{clauser1969proposed},
 \begin{equation}
  p\left(\varphi_{1},\varphi_{2}\right)=
  \cos^{2}{\left(\varphi_{1}-\varphi_{2}\right)}.
 \end{equation}
 The difference between this approach and the assumptions made in Bell's 
 inequalities is that the pseudoscalar terms in the rotors have an effect on 
 the correlation probabilities. Thus, we find that measurements conducted on 
 the same axis are expected to be fully correlated, whereas perpendicular 
 measurements will be uncorrelated. This correlation is explicitly contained 
 in local variables, unlike the phase correlations proposed by de 
 Broglie~\cite{de1927wave}, and later Bohm~\cite{bohm1952suggested}, which are 
 manifestly non-local. In the current model, the superluminal communication is 
 simply an artefact of the phase correlation and, since this correlation does 
 not violate local causality, then there is no paradox.  
 
\section{Conclusion}
 In conclusion, within the standard approach, it is assumed that electron spin 
 cannot have the properties of a vector and still maintain its isotropy. 
 However, in order to interact with a magnetic field, electron spin \emph{must} 
 have the properties of a vector. We assert this conflict is the source of 
 many of the paradoxes related to electron spin and that exploiting a model in 
 which the spin is described consistently in real three-dimensional space 
 allows us to resolve these paradoxes while maintaining the isotropy. 
 
 The essential difference between this model and conventional interpretations 
 is that the electrons are modelled as spatially extended entities as opposed 
 to point-particles. In this way, the wave properties are encoded into 
 oscillating mass and spin densities, which comply with the Born rule to give 
 the inertial number density. Spin-up and spin-down are represented by spin 
 vectors that are respectively parallel and antiparallel to the vector of 
 motion of the electron, which is itself an extended vector field. The 
 isotropy of the electron spin is reproduced in a statistical manifold with an 
 equal number of spin-up and spin-down electrons. Moreover, this behaviour is 
 a manifestation of the helicity of electromagnetic field components, the 
 orientation of which may be affected by an external magnetic field, giving 
 rise to the results of Stern-Gerlach-type experiments. In principle, this 
 process is deterministic, since the spin density determines the result. In 
 practice, the spin density is unknown and the experimental results must still 
 be analysed statistically.
 
 We have also shown that the non-commutativity of electron spin measurements 
 on different axes is well explained by the interaction between the spin 
 vectors and the measurement field. Finally, EPR-type experiments were 
 interpreted through the lens of extended electrons and the spectre of spooky 
 action at a distance was found to be nothing more than an underlying 
 correlation. 
\begin{acknowledgments}
 The authors acknowledge EPSRC funding for the UKCP consortium, grant 
 No. EP/K013610/1
\end{acknowledgments}


\begin{thebibliography}{10}

\bibitem{jackson1999classical}
John~David Jackson.
\newblock {\em {Classical electrodynamics}}.
\newblock Wiley, 1999.

\bibitem{eisberg1986quantum}
Robert Eisberg, Robert Resnick, and Judith Brown.
\newblock {Quantum physics of atoms, molecules, solids, nuclei, and particles}.
\newblock {\em Physics Today}, 39:110, 1986.

\bibitem{mahajan1988electricity}
Anant~S Mahajan and Abbas~A Rangwala.
\newblock {\em {Electricity and Magnetism}}.
\newblock Tata McGraw-Hill Education, 1988.

\bibitem{gerlach1922experimentelle}
Walther Gerlach and Otto Stern.
\newblock {Der experimentelle nachweis der richtungsquantelung im magnetfeld}.
\newblock {\em Zeitschrift f{\"u}r Physik A Hadrons and Nuclei}, 9(1):349--352,
  1922.

\bibitem{burkhardt2008foundations}
Charles~E Burkhardt and Jacob~J Leventhal.
\newblock {\em {Foundations of quantum physics}}.
\newblock Springer Science \& Business Media, 2008.

\bibitem{rieder2004scanning}
Karl-Heinz Rieder, Gerhard Meyer, Saw-Wai Hla, Francesca Moresco, Kai~F Braun,
  Karina Morgenstern, Jascha Repp, Stefan Foelsch, and Ludwig Bartels.
\newblock {The scanning tunnelling microscope as an operative tool: doing
  physics and chemistry with single atoms and molecules}.
\newblock {\em Philosophical Transactions of the Royal Society of London A:
  Mathematical, Physical and Engineering Sciences}, 362(1819):1207--1216, 2004.

\bibitem{hofer2012heisenberg}
Werner~A Hofer.
\newblock {Heisenberg, uncertainty, and the scanning tunneling microscope}.
\newblock {\em Frontiers of Physics}, 7(2):218--222, 2012.

\bibitem{hofer2011unconventional}
Werner~A Hofer.
\newblock {Unconventional approach to orbital-free density functional theory
  derived from a model of extended electrons}.
\newblock {\em Foundations of Physics}, 41(4):754--791, 2011.

\bibitem{hofer2014elements}
Werner~A Hofer.
\newblock {Elements of physics for the 21st century}.
\newblock In {\em {Journal of Physics: Conference Series}}, volume 504, page
  012014. IOP Publishing, 2014.

\bibitem{hestenes2012clifford}
David Hestenes and Garret Sobczyk.
\newblock {\em {Clifford Algebra to Geometric Calculus: A unified language for
  mathematics and physics}}, volume~5.
\newblock Springer Science \& Business Media, 2012.

\bibitem{gull1993imaginary}
Stephen Gull, Anthony Lasenby, and Chris Doran.
\newblock {Imaginary numbers are not real: the geometric algebra of spacetime}.
\newblock {\em Foundations of Physics}, 23(9):1175--1201, 1993.

\bibitem{benenti2004principles}
Giuliano Benenti, Giuliano Strini, and Giulio Casati.
\newblock {\em {Principles of quantum computation and information}}.
\newblock World scientific, 2004.

\bibitem{ghirardi1986unified}
Gian~Carlo Ghirardi, Alberto Rimini, and Tullio Weber.
\newblock {Unified dynamics for microscopic and macroscopic systems}.
\newblock {\em Physical Review D}, 34(2):470, 1986.

\bibitem{penrose1996gravity}
Roger Penrose.
\newblock {On gravity's role in quantum state reduction}.
\newblock {\em General relativity and gravitation}, 28(5):581--600, 1996.

\bibitem{heisenberg1958language}
Werner Heisenberg.
\newblock {\em {Language and reality in modern physics}}.
\newblock na, 1958.

\bibitem{knee2016strict}
George~C Knee, Kosuke Kakuyanagi, Mao-Chuang Yeh, Yuichiro Matsuzaki, Hiraku
  Toida, Hiroshi Yamaguchi, Anthony~J Leggett, and William~J Munro.
\newblock {A strict experimental test of macroscopic realism in a
  superconducting flux qubit}.
\newblock {\em arXiv preprint arXiv:1601.03728}, 2016.

\bibitem{debroglie1925research}
L~{De Broglie}.
\newblock {Research on the theory of quanta}.
\newblock In {\em {Annales de Physique}}, volume~10, pages 22--128, 1925.

\bibitem{schrodinger1926undulatory}
Erwin Schr{\"o}dinger.
\newblock {An undulatory theory of the mechanics of atoms and molecules}.
\newblock {\em Physical Review}, 28(6):1049, 1926.

\bibitem{bohm1952suggested}
David Bohm.
\newblock {A suggested interpretation of the quantum theory in terms of
  "hidden" variables. I}.
\newblock {\em Physical Review}, 85(2):166, 1952.

\bibitem{bohm1952suggestedII}
David Bohm.
\newblock {A suggested interpretation of the quantum theory in terms of
  "hidden" variables. II}.
\newblock {\em Physical Review}, 85(2):180, 1952.

\bibitem{bell1966problem}
John~S Bell.
\newblock {On the problem of hidden variables in quantum mechanics}.
\newblock {\em Reviews of Modern Physics}, 38(3):447, 1966.

\bibitem{everett1957relative}
Hugh {Everett III}.
\newblock {" Relative state" formulation of quantum mechanics}.
\newblock {\em Reviews of Modern Physics}, 29(3):454, 1957.

\bibitem{hooft2007free}
Gerard't Hooft.
\newblock {The free-will postulate in quantum mechanics}.
\newblock {\em arXiv preprint quant-ph/0701097}, 2007.

\bibitem{hooft2009entangled}
Gerard't Hooft.
\newblock {Entangled quantum states in a local deterministic theory}.
\newblock {\em arXiv preprint arXiv:0908.3408}, 2009.

\bibitem{de1976time}
O~Costa de~Beauregard.
\newblock {Time symmetry and interpretation of quantum mechanics}.
\newblock {\em Foundations of Physics}, 6(5):539--559, 1976.

\bibitem{dowe1997defense}
Phil Dowe.
\newblock {A defense of backwards in time causation models in quantum
  mechanics}.
\newblock {\em Synthese}, 112(2):233--246, 1997.

\bibitem{santos2004failure}
Emilio Santos.
\newblock {The failure to perform a loophole-free test of Bell{\rq}s 
inequality supports local realism}.
\newblock {\em Foundations of Physics}, 34(11):1643--1673, 2004.

\bibitem{brunner2014bell}
Nicolas Brunner, Daniel Cavalcanti, Stefano Pironio, Valerio Scarani, and
  Stephanie Wehner.
\newblock {Bell nonlocality}.
\newblock {\em Reviews of Modern Physics}, 86(2):419, 2014.

\bibitem{hofer2012solving}
Werner~A Hofer.
\newblock {Solving the Einstein-Podolsky-Rosen puzzle: the origin of
  non-locality in Aspect-type experiments}.
\newblock {\em Frontiers of Physics}, 7(5):504--508, 2012.

\bibitem{doran1993states}
Chris Doran, Anthony Lasenby, and Stephen Gull.
\newblock {States and operators in the spacetime algebra}.
\newblock {\em Foundations of physics}, 23(9):1239--1264, 1993.

\bibitem{einstein1935can}
Albert Einstein, Boris Podolsky, and Nathan Rosen.
\newblock {Can quantum-mechanical description of physical reality be considered
  complete?}
\newblock {\em Physical review}, 47(10):777, 1935.

\bibitem{einstein1936physics}
Albert Einstein.
\newblock {Physics and reality}.
\newblock {\em Journal of the Franklin Institute}, 221(3):349--382, 1936.

\bibitem{thaller2005advanced}
Bernd Thaller.
\newblock {\em {Advanced visual quantum mechanics}}.
\newblock Springer Science \& Business Media, 2005.

\bibitem{wittmann2012loophole}
Bernhard Wittmann, Sven Ramelow, Fabian Steinlechner, Nathan~K Langford,
  Nicolas Brunner, Howard~M Wiseman, Rupert Ursin, and Anton Zeilinger.
\newblock {Loophole-free Einstein--Podolsky--Rosen experiment via quantum
  steering}.
\newblock {\em New Journal of Physics}, 14(5):053030, 2012.

\bibitem{clauser1969proposed}
John~F Clauser, Michael~A Horne, Abner Shimony, and Richard~A Holt.
\newblock {Proposed experiment to test local hidden-variable theories}.
\newblock {\em Physical review letters}, 23(15):880, 1969.

\bibitem{de1927wave}
Louis de~Broglie.
\newblock {Wave mechanics and the atomic structure of matter and of radiation}.
\newblock {\em Le Journal de Physique et le Radium}, 8:225, 1927.

\end{thebibliography}
\end{document}